\documentclass[onecolumn,aps,nofootinbib,preprint]{revtex4}

\usepackage{graphicx}
\usepackage{amsmath}
\usepackage{hyperref}
\usepackage{amssymb}
\usepackage{color}

\newcommand{\beq}{\begin{equation}}
\newcommand{\eeq}{\end{equation}}
\newcommand{\bea}{\begin{eqnarray}}
\newcommand{\eea}{\end{eqnarray}}

\newcommand{\SUw}{\ensuremath{SU(2)_{weak}\ }}
\newcommand{\SUN}{\ensuremath{SU(N)_d \ }}

\newcommand{\ignore}[1]{}

\makeatletter
\def\simgt{\mathrel{\lower2.5pt\vbox{\lineskip=0.5pt\baselineskip=0pt
           \hbox{$>$}\hbox{$\sim$}}}}
\def\simlt{\mathrel{\lower2.5pt\vbox{\lineskip=0.5pt\baselineskip=0pt
           \hbox{$<$}\hbox{$\sim$}}}}
\makeatother

\begin{document}
\setlength{\unitlength}{1mm}

\title{Non-Abelian dark matter and dark radiation}

\author{Manuel A. Buen-Abad}\email{buenabad@bu.edu}
\author{Gustavo Marques-Tavares}\email{gusmt@bu.edu}
\author{Martin Schmaltz}\email{schmaltz@bu.edu}
\affiliation{Physics Department, Boston University \\ 590 Commonwealth Avenue \\ Boston, MA 02215 \\ USA  \\ \ }

\begin{abstract}
\vskip.1in
We propose a new class of dark matter models with unusual phenomenology. What is ordinary about our models is that dark matter particles are WIMPs, they are weakly coupled to the Standard Model and have weak scale masses. What is unusual is that they come in multiplets of a new ``dark" non-Abelian gauge group with milli-weak coupling. The massless dark gluons of this dark gauge group contribute to the energy density of the universe as a form of weakly self-interacting dark radiation. 
In this paper we explore the consequences of having i.) dark matter in multiplets ii.) self-interacting dark radiation and iii.) dark matter which is weakly coupled to dark radiation.  
We find that i.) dark matter cross sections are modified by multiplicity factors which have significant consequences for collider searches and indirect detection, ii.) dark gluons have thermal abundances which affect the CMB as dark radiation. Unlike additional massless neutrino species the dark gluons are interacting and have vanishing viscosity and iii.) the coupling of dark radiation to dark matter represents a new mechanism for damping the large scale structure power spectrum. A combination of additional radiation and slightly damped structure is interesting because it can remove tensions between global $\Lambda$CDM fits from the CMB and direct measurements of the Hubble expansion rate ($H_0$) and large scale structure ($\sigma_8$).

\end{abstract}

\maketitle

\section{Introduction}

Dark matter (DM) makes up more than $80 \%$ of the matter content of our universe \cite{Planck:2015xua}. Despite the overwhelming quantitative evidence for cold dark matter (CDM) from its gravitational interactions, and its inclusion in the standard $\Lambda$CDM cosmological model, we only have very limited information about possible non-gravitational interactions of DM.
Existing upper bounds on interactions between the DM and the Standard Model (SM) and self-interactions of the DM leave a broad range of possibilities for models of DM. Viable models may include a whole dark sector consisting of different species of particles which interact only with themselves and with DM~(see e.g. \cite{Feng:2008mu,Feng:2009mn,Franca:2013zxa}). Furthermore, there is some tension 
between the best fit parameters in $\Lambda$CDM as inferred from the cosmic microwave background (CMB)~\cite{Planck:2015xua}, and direct measurements of $H_0$ using standard candles and $\sigma_8$  using galaxy clusters~\cite{Riess:2011yx,Beutler:2014yhv,Battye:2014qga}). While it is premature to draw definite conclusions about these discrepancies, they motivate the study of interacting dark sectors which go beyond the ordinary {cold dark matter} of $\Lambda$CDM.

In this work, we propose a novel scenario for an interacting dark sector which predicts several important modifications to standard CDM phenomenology. Our dark sector includes a weakly interacting massive particle (WIMP) as the dark matter. As usual, weak interactions between the DM and the SM keep the two sectors in equilibrium at high temperatures and set the abundance of DM through thermal freeze-out. The non-trivial new ingredient of our scenario is that the DM particles are charged under a new non-Abelian ``dark" gauge group.\footnote{%
New visible particles charged under a new non-Abelian gauge group with macroscopic sized confinement scale have been proposed in the past under the name of theta-particles~\cite{Okun:1980mu,Okun:1980kw,Khlopov:1980ar} and more recently as quirks~\cite{Kang:2008ea}.
} Thus our dark sector consists of a degenerate multiplet of massive DM particles coupled to ``dark radiation" in the form of massless non-Abelian gauge bosons. Observations of large scale structure tightly constrain long-range interactions of DM with dark radiation and require very small gauge couplings\footnote{
Such couplings are too small to have appreciable effects on the Bullet Cluster, the shape of galactic dark matter halos or DM protohalos. They are also too small to have any bearing on the ``core vs cusp" or ``missing satellites" problems.}, 
$g_d < 10^{-3}$, so that the ``dark" confinement temperature is well below the current  CMB temperature.\footnote{
For recent discussions of non-Abelian dark sectors in a different context see~\cite{Feng:2011ik,Jeong:2013eza,Baek:2013dwa,Boddy:2014yra,Blinov:2014nla,Yamanaka:2014pva}.}
Thus confinement in the dark sector is irrelevant, and the ``dark gluons" are a thermal bath of weakly coupled massless particles which are well described as a perfect fluid. Note that this is distinct from neutrinos which free-stream and are therefore not a perfect fluid. The two types of dark radiation can be distinguished through their imprints on the CMB~\cite{Friedland:2007vv,Diamanti:2012tg}.

The general features of our dark sector lead to several distinct observable effects which can be broadly classified as due to ({\it i.}) multiplicity of the dark matter, ({\it ii.}) dark radiation in the form of the dark gluon fluid, and ({\it iii.}) interactions between DM and dark radiation. We briefly summarize the salient features of each of the three effects here. Details will be provided in subsequent sections. 

\begin{itemize}

\item {\it DM multiplicity:} Being charged under an unbroken non-Abelian gauge group, DM particles come in degenerate multiplets of $N$ dark colors. This multiplicity leads to several easy-to-understand but important differences from conventional WIMPs. First off, DM annihilation cross sections (into SM particles) are suppressed by $1/N$ from averaging over initial states. Assuming thermal freeze-out, we predict that our dark matter mass is smaller by a factor of $1/\sqrt{N}$ than in models without dark matter multiplicity. Similarly, indirect detection bounds from dark matter annihilation into photons near the galactic center are relaxed because of the $1/N$ in the annihilation cross section.  Second, DM pair production cross sections at colliders are enhanced by a factor of $N$ from the multiplicity of possible final states. Third, direct detection bounds for DM are unchanged because the scattering cross section for DM particles off nuclei does not contain multiplicity factors. We discuss these effects, which would also follow from a non-Abelian global symmetry, in Section 3.

\item {\it Dark radiation:} Through their interactions with the DM, dark gluons come into thermal equilibrium with the SM in the early universe. After freeze-out of the DM the dark gluons decouple from the SM. They maintain a thermal distribution but end up cooling relative to the CMB  because the photons are heated by absorbing the entropy contained in massive SM particles. The dark gluons are observable through their contribution to the radiation density in the universe, an effect which is conventionally expressed in terms of an effective number of neutrino species~(see \cite{Brust:2013ova} for a recent discussion of bounds on new relativistic degrees of freedom from the CMB). We find $\triangle N_{eff} \sim 0.07 (N^2-1)$ for the $N^2-1$ gluons of an \SUN gauge group. Bounds on $N_{eff}$ from Planck and from nucleosynthesis give $N<4$ at 95\% confidence level. An increase in the radiation density also shifts the best fit values of other cosmological parameters from the CMB. The two most significant shifts are increases in the predicted values of $H_0$ and $\sigma_8$. An increase of $H_0$ would remove the tension between CMB fits and direct measurements of $H_0$. An increase in $\sigma_8$ worsens the tension between CMB data and direct observations of $\sigma_8$ from large scale structure. However, as discussed in the next item, interactions between DM and dark gluons can remove this tension. Finally, the fact that dark gluons are self-interacting allows one to distinguish them from additional neutrino species. Dark gluons are well described by a perfect fluid with no viscosity and therefore lead to less damping of the CMB power spectrum in comparison to extra free-streaming relativistic fluids. Future CMB data will be precise enough to determine the viscosity of any significant new component of dark radiation and can therefore distinguish between additional neutrinos and dark gluons~\cite{Diamanti:2012tg}. We elaborate on the phenomenology of our dark radiation in Section 4.

\item {\it DM - dark radiation interactions:} Scattering of dark radiation (DR) off DM introduces a drag force between  the non-relativistic DM fluid and the relativistic radiation. This drag suppresses gravitational clustering and is therefore observable in the matter power spectrum. Allowing for matter perturbations to grow at least approximately as in $\Lambda$CDM gives a conservative upper bound of $\alpha_d \simlt 10^{-8}$. What we find particularly interesting about our scenario is that the momentum transfer cross section for DM - DR scales with temperature like the Hubble parameter during radiation domination. This means that the effect of the radiation remains equally important throughout radiation domination, leading to a smooth reduction of the power spectrum. This is in contrast with the more frequently studied case where the DM interactions freeze out at a critical temperature, leading to a sharp cutoff in the power spectrum at small scales~\cite{Hofmann:2001bi,Loeb:2005pm,Bertschinger:2006nq,Feng:2009mn}. We show that a very small and smooth reduction of the power spectrum may resolve the tension in the indirect determination of $\sigma_8$ from Planck data and direct measurements of the power spectrum at large scales~(see e.g. \cite{Planck:2015xua,Beutler:2014yhv,Battye:2014qga}).\footnote{%
We thank J. Lesgourgues for alerting us to the discrepancy between determinations of $\sigma_8$ from fits to the CMB and from direct measurements.}
We discuss DM-DR interactions in Section 5.

\end{itemize}

Most of the features associated with non-Abelian dark matter are expected to hold in a broad class of models irrespective of the specific interactions between the DM and the SM. However,  in the interest of concreteness and for clarity of presentation we focus on a specific example model which is described in Section 2. The DM in the model is a ``wino-like" \SUw triplet Dirac fermion which transforms as a fundamental under a dark \SUN gauge group.

\section{The Model and DM thermal relic abundance}

For concreteness we focus on a specific realization of a non-Abelian dark matter. We take the dark matter particle to be a Dirac fermion in the $(1,3)_0$ representation of the Standard Model $(SU(3),SU(2))_{U(1)}$ gauge group and in the fundamental representation of dark \SUN. It has an \SUw preserving Dirac mass $M_\chi$ and no additional interactions with the Standard Model besides gauge interactions. Electroweak symmetry breaking leads to a loop-induced mass splitting between the charged and neutral components of the \SUw triplet. In the limit $M_\chi \gg M_W$ the splitting is independent of $M_\chi$ and given by~\cite{Cheng:1998hc,Ibe:2012sx}
\beq
\delta M_\chi = M_{\chi^{\pm}} - M_{\chi^{0}} \approx 0.16 \, \text{GeV} \, .
\label{eq:mass-splitting}
\eeq
The neutral component $\chi^0$ is the lightest particle which carries fundamental \SUN charge and is therefore stable, it is our DM candidate. From the perspective of SM interactions the dark \SUN is simply a global symmetry, so that our DM acts like $N$ identical copies of a weakly interacting particle $\chi^0$ with Dirac mass $M_\chi$. In the following, we determine $M_\chi$ by requiring that the correct DM abundance is obtained from thermal freeze-out.  

At the time of DM chemical freeze-out $T\sim M_\chi/26 \gg \delta M_\chi$ so that we can ignore the mass splitting between the different components of $\chi^a$ when calculating relic abundances. Once the temperature drops below $\delta M_\chi$, the charged $\chi^\pm$ decay to $\chi^0$ plus standard model particles. Thus the (co-moving) number density of $\chi^0$ particles at low temperatures is simply given by summing over all components of $\chi^a$ at DM freeze-out, i.e.  $\sum_a n_{\chi^a} \rightarrow  n_{\chi^0}$. This explains why we can use the abundance calculation for the whole triplet $\chi^a$ in order to find the DM relic abundance.

\begin{figure}[ht]
\centering
\includegraphics[width=0.8\textwidth]{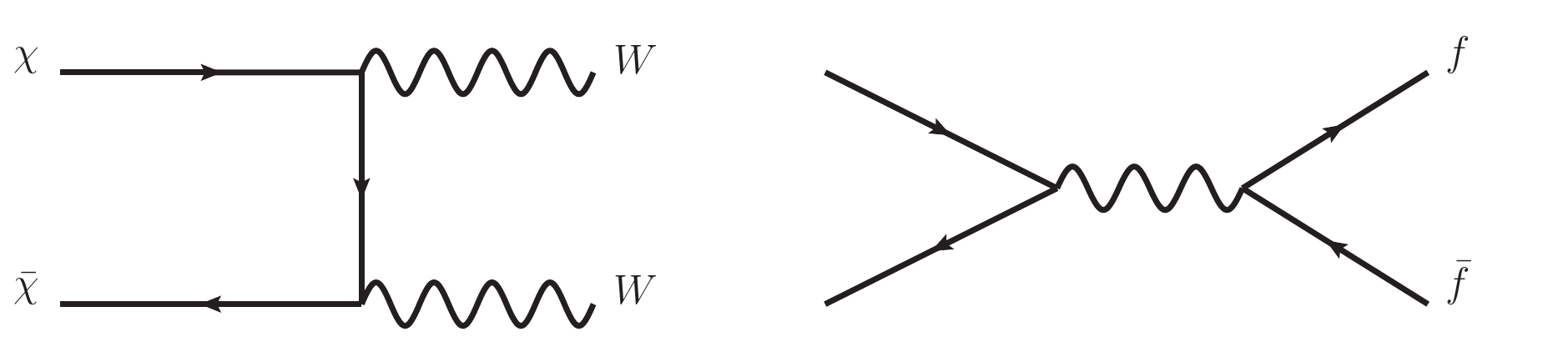}
\caption{Annihilation of dark matter into \SUw gauge bosons or SM fermions. Since the mass splitting between members of \SUw multiplets is small compared with energy transfer in the annihilation diagrams (i.e. twice the $\chi$ mass) we compute the co-annihilation of full \SUw multiplets and ignore the mass splittings.}
\label{fig:annihilation}
\end{figure}

We will show in subsequent sections that Cosmology requires the dark gauge coupling to be much smaller than the SM gauge couplings. Therefore the relevant interactions for the thermal relic calculations are of the form $\chi \chi \rightarrow \text{SM SM}$ and independent of the dark gauge coupling. Figure~\ref{fig:annihilation}  shows some of the relevant diagrams for DM annihilating to SM particles. In the limit $M_\chi^2 \gg M_W^2$, the thermally averaged effective annihilation cross-section is
\beq
\langle \sigma v \rangle = \frac{1}{2 N} \, \frac{37 g_2^4}{192 \pi M_\chi^2} \, ,
\label{eq:annihilation-xsec}
\eeq
where $g_2$ is the \SUw gauge coupling. This cross-section differs from the standard \SUw triplet ``wino"~\cite{Cirelli:2005uq} by the extra $1/2N$ factor, which comes from the multiplicity associated with the \SUN representation and the fact that $\chi$ is a 4-component Dirac fermion instead of a 2-component Majorana fermion. The $1/2N$ factor can be easily understood from the fact that any given DM particle carries an dark color charge and can only annihilate if it finds an anti-particle with the corresponding anti-dark color, thus reducing the color averaged annihilation cross-section.

If the DM abundance is set by thermal freeze-out, then - to a good approximation - the DM mass density today depends on the DM mass only through its annihilation cross-section~\cite{Kolb:1990vq}. Therefore, holding the cross section in Eq.~\ref{eq:annihilation-xsec} fixed at the Cosmologically preferred value we see that the mass required to get the correct relic abundance decreases as the square root of $N$. In Table~\ref{table:masses} we give the mass of DM for different values of $N$ using the tree level annihilation cross-section. The masses are significantly lower than for the usual ``wino" case where the tree level formula predicts a mass of $2.4$ TeV.~\footnote{In the case of the ``wino", the preferred mass is close to a Sommerfeld resonance of the weak interactions and one must account for Sommerfeld enhancement to obtain $M_{wino} \simeq 3$ TeV ~\cite{Cirelli:2007xd}. In our case, the DM mass is safely below the Sommerfeld resonances and the perturbative annihilation formula \ref{eq:annihilation-xsec} is adequate.}

\begin{table}[ht]
\centering
\begin{tabular}{|c|c|c|c|c|}
\hline
$N=2$ & $N=3$ & $N=4$ & $N=5$ & Generic $N$ \\ 
\hline \hline
$1.2$ TeV & $1.0$ TeV & $0.9$ TeV & $0.8$ TeV & $\sim {2.4}/{\sqrt{2N}}$ TeV  \\
\hline
\end{tabular}
\caption{Dark Matter masses required to get the correct thermal abundance as a function of $N$.}
\label{table:masses}
\end{table}

\section{Dark Matter multiplicity and experimental searches}

As will be discussed in subsequent Sections, astrophysical and cosmological constraints require very small dark gauge couplings of order $\alpha_d < 10^{-8}$. This is too small to have observable effects for production of DM at colliders, or for direct or indirect detection of DM. However, it is important to take into account that DM particles are in multiplets of the dark \SUN, i.e. there are multiplicity factors associated with DM processes. In the previous Section we already discussed that the predicted mass for our thermal relic DM is reduced by a multiplicity factor of $1/\sqrt{2N}$, which can lead to large changes in sensitivity of experimental searches. Similar multiplicity/color factors also appear in cross-sections associated with DM detection. Figure~\ref{fig:searches} shows the color factors associated with the different kinds of DM searches. In what follows we briefly describe the effects of \SUN multiplicity for direct and indirect detection and also for collider searches for DM:

\begin{figure}[h!t]
\centering
\includegraphics[width=0.5\textwidth]{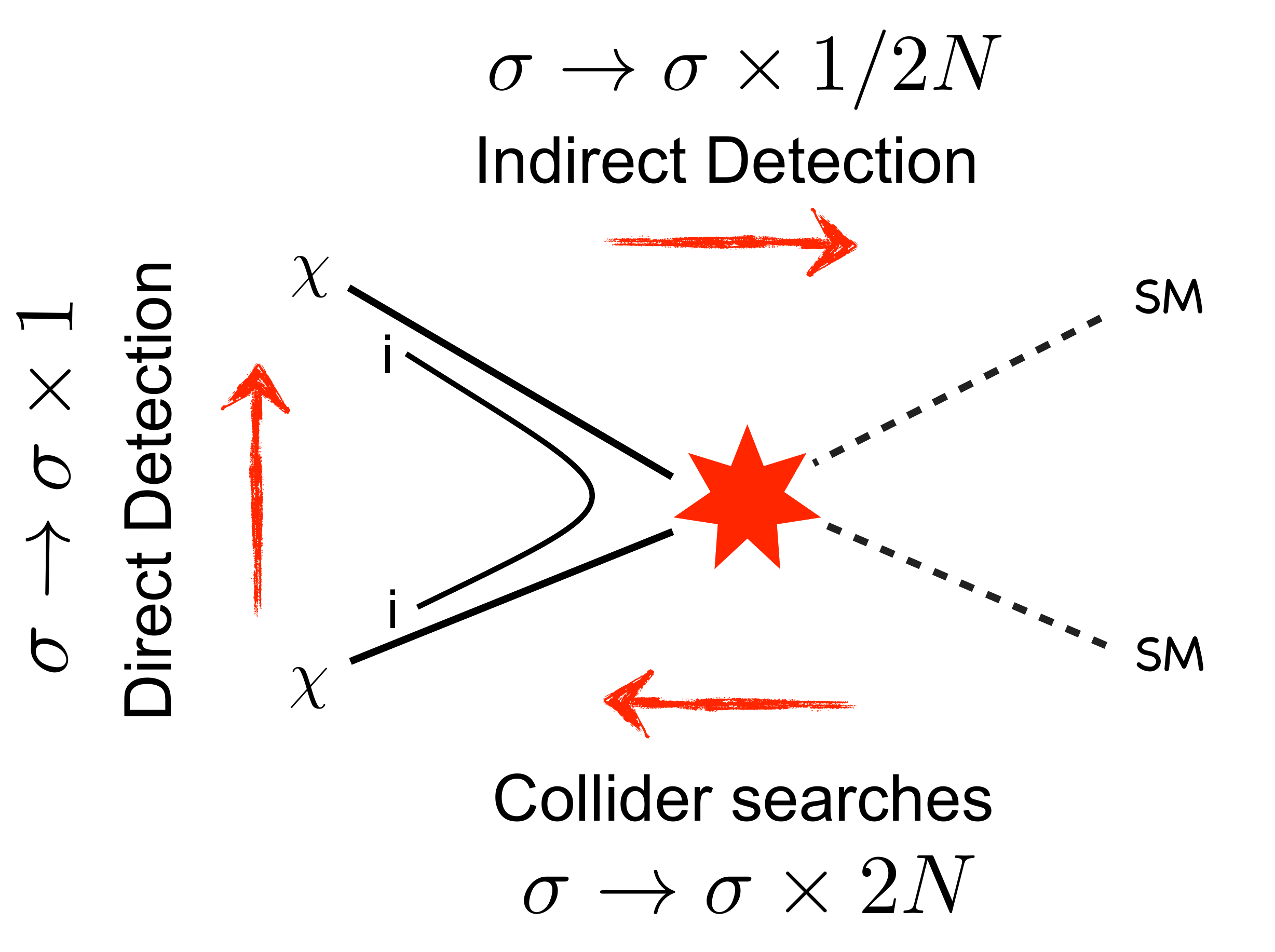}
\caption{Color factors for the different types of DM search experiments. The different multiplicity factors can be easily understood from the color flow in the figure. For direct detection the color of the incoming dark matter is the same as of the outgoing and so there is no multiplicity factor. For indirect detection it is an annihilation diagram, so just as for the thermal relic calculation there is a $1/2N$ suppression because a DM particle can only annihilate if it finds the anti-particle with the right anti-dark color. For colliders there is an $2N$ enhancement because any of the different N colors can be created and an extra 2 from Dirac vs Majorana.}
\label{fig:searches}
\end{figure}

\subsection{Direct detection}

 For direct detection there is no color factor associated with the multiplicity, thus the only change comes from dark matter being lighter. The spin-independent cross-section for dark matter scattering of the nucleus is approximately $10^{-47} \, \text{cm}^2$, and independent of the DM mass as long as $M_\chi \gg M_W$~\cite{Hill:2013hoa}. This cross-section is an order of magnitude smaller than the projected sensitivity of the next generation direct detection experiments~\cite{Cushman:2013zza}. However, in the mass range of interest (around $1 \, \text{TeV}$) it is above the neutrino background and potentially within reach of future experiments.

\subsection{Indirect detection}

\begin{figure}[h!t]
\centering
\includegraphics[width=0.6\textwidth]{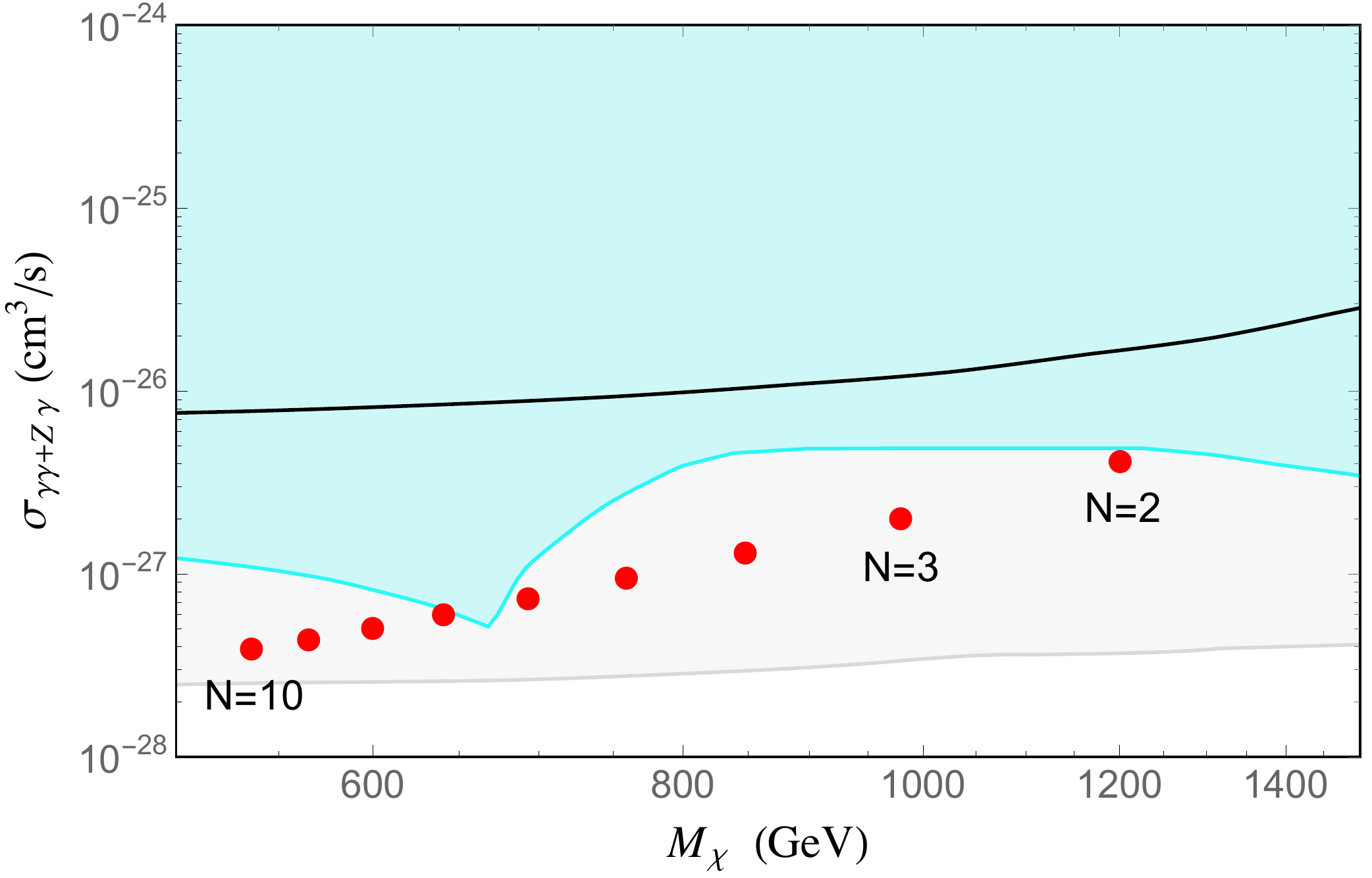}
\caption{Indirect detection constraints from gamma ray line searches from H.E.S.S. (blue region) and projected sensitivity for gamma ray lines at CTA (gray region) assuming an NFW profile for the dark matter distribution in the galactic center, both taken from~\cite{Ovanesyan:2014fwa}. The red dots are the cross sections for dark matter annihilation into gamma rays in our models for $N=2$ up to $N=10$, which were obtained by appropriately rescaling the NLL cross section with Sommerfeld enhancement from~\cite{Ovanesyan:2014fwa}. For comparison we plotted in black the cross-section for the annihilation cross-section to photons in the standard ``wino'' model as a function of mass, also from~\cite{Ovanesyan:2014fwa}.}
\label{fig:indirect}
\end{figure}

The annihilation cross-section relevant for indirect detection is suppressed by a $1/2N$ factor. Taking into account Sommerfeld enhancement the cross-section is further reduced relative to the standard ``wino" model. This is because there is a resonant Sommerfeld enhancement from weak interactions for dark matter masses in the $2-3 \, \text{TeV}$ range, but the enhancement is much smaller for masses around $1 \, \text{TeV}$. The ``wino" model has been investigated recently~\cite{Cohen:2013ama,Fan:2013faa,Bauer:2014ula,Ovanesyan:2014fwa,Baumgart:2014saa}, and is strongly disfavored by H.E.S.S. data. Our \SUN model is not yet constrained by either H.E.S.S. or Fermi data for any $N\ge 2$, but the annihilation cross-section is close to H.E.S.S. sensitivity as shown in Figure~\ref{fig:indirect}. It is worth noting that the limits shown in Figure~ \ref{fig:indirect} assume a specific NFW profile and there is a large uncertainty in these limits due to our limited knowledge of the dark matter distribution in the center of the galaxy. Also shown in the figure is the projected reach of CTA~\cite{Funk:2013gxa}, assuming 5 hours of observation time. One can see that CTA should have enough sensitivity to discover or rule out our model up to at least $N=10$, and therefore is the most promising search for discovering dark matter in our model.

\subsection{LHC and future collider searches}

The multiplicity factor enhances sensitivity of collider searches to our DM in two ways. Most importantly, the predicted DM mass from thermal freeze-out is lowered,  and thus a lower partonic center of mass energy is required to pair produce DM at a collider. For example,  at the 14 TeV LHC and for DM masses near $M_{DM} \sim$ 500 GeV the DM cross section scales as $(1/M_{DM})^6$ because of the strong energy dependence of the parton luminosities. In addition, the cross-section for pair producing dark matter is enhanced by the final state multiplicity factor $2N$.

\begin{figure}[h!t]
\centering
\includegraphics[width=0.5\textwidth]{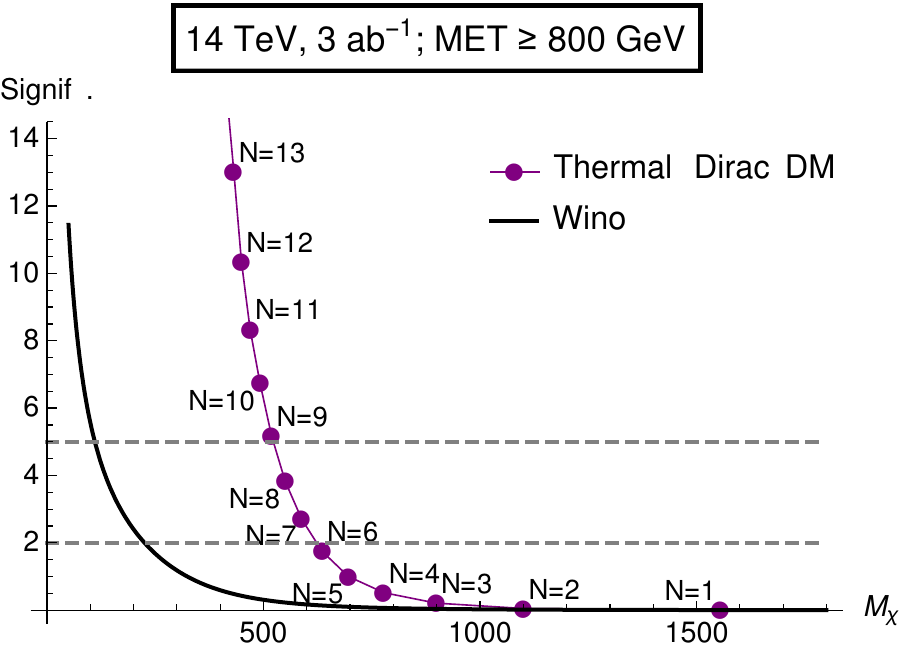}
\hskip-5pt
\includegraphics[width=0.5\textwidth]{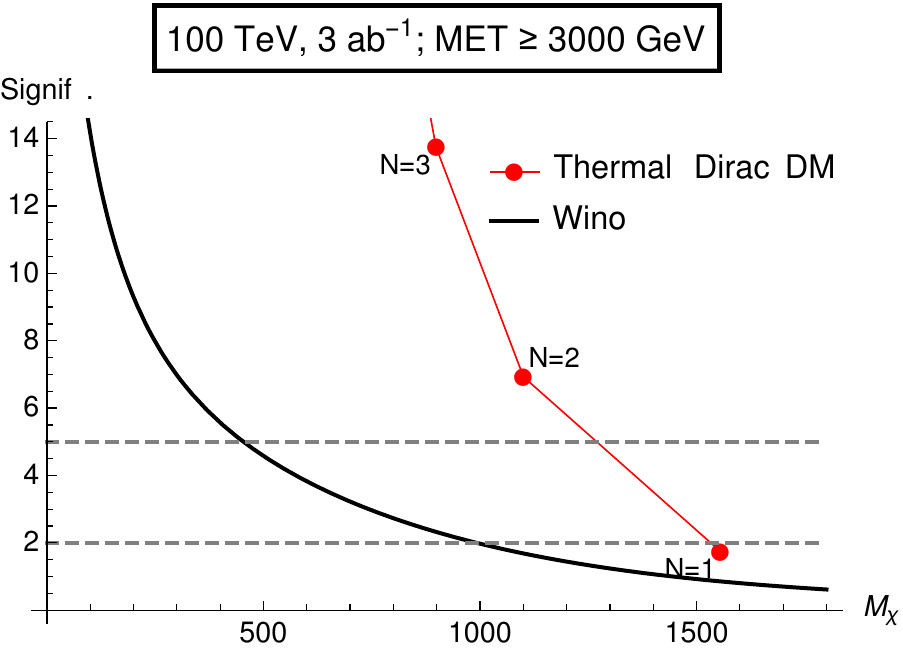}
\caption{Expected significance of missing energy (MET) searches for DM at the LHC and a future 100 TeV collider. The solid black lines in each plot correspond to the sensitivity of the collider to ``wino"-like DM, an \SUw triplet Majorana fermion. The colored dots labeled by different $N$-values correspond to our models in which the DM is a Dirac fermion with multiplicity $N$ and mass chosen to yield the correct abundance from thermal freeze-out.}
\label{fig:LHC}
\end{figure}

In the left panel of Figure~\ref{fig:LHC} we show the expected sensitivity of the high luminosity LHC to the DM in our model. The solid dots correspond to DM with multiplicity $N$ and Dirac masses chosen so that the correct thermal DM abundance is obtained. We see that the 14 TeV LHC is sensitive to DM with $N\ge7$. For comparison, we also show the expected sensitivity to a standard ``wino" \SUw triplet with a Majorana mass. Note that the ``wino" has the correct thermal abundance only for $M_{wino}\simeq 3$ TeV. Existing monojet searches from ATLAS~\cite{Aad:2015zva} and CMS~\cite{Khachatryan:2014rra} with 8 TeV collisions and a luminosity of 19.5 fb$^{-1}$ already rule out $N\simgt 20$.

In the right panel we show that a 100 TeV future collider can discover our DM above backgrounds even for the smallest non-Abelian multiplet of $N=2$ and perhaps may be able to rule out a Dirac ``wino". The significances for these plots were determined from parton level signal and background events which were computed with MadGraph \cite{Alwall:2011uj}. The main irreducible backgrounds are due to jets plus $Z$ or $W$ with MET from decays to neutrinos. We computed signal and backgrounds to leading order ($\alpha_s \alpha_W$) and assumed that the experiments will be able to limit background systematic uncertainties to 2\%. For the ``wino", our results are consistent with the more sophisticated studies in (\cite{Low:2014cba,Cirelli:2014dsa}). These references also showed that a MET plus ``disappearing track" search for the production of $\chi^\pm$ can improve sensitivity because it is free of irreducible SM backgrounds. 

Another observable consequence of this model is a change in the running of the EW gauge coupling~\cite{Alves:2014cda}. The multiplicity of DM leads to an $2N$ enhancement factor in the DM contribution to running of $\alpha_W$ at one loop, which would be observable at the proposed 100 TeV hadron colliders.

\section{Dark gluons as dark radiation}

In this Section we turn our attention to the cosmic evolution of the energy density in dark gluons and its effects on the CMB. An important parameter which determines the effects of dark gluons on the CMB is the ratio of temperatures of the dark gluon plasma $T_d$ over the photon temperature $T$.

\begin{figure}[ht]
\centering
\includegraphics[width=0.5\textwidth]{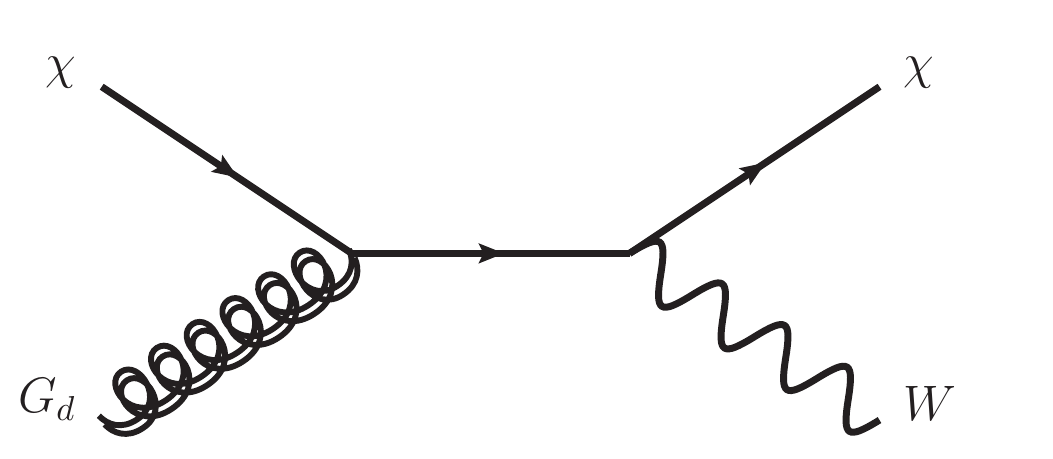}
\caption{The process through which dark gluons maintain equilibrium with the DM and the SM plasma for small $\alpha_d$.}
\label{fig:darkgluon-compton}
\end{figure}

Dark gluons are coupled to the thermal bath of SM particles through their couplings to the DM which is in equilibrium with the SM in the early universe. If $\alpha_d$ is not too small the dark gluons also equilibrate with the SM plasma at early times, so that $T_d=T$ before DM freeze-out. To determine the smallest possible coupling $\alpha_d$ for which the dark gluons are in chemical and kinetic equilibrium with the SM we consider the Feynman diagram in Figure~\ref{fig:darkgluon-compton}. At temperatures higher than $M_\chi$, the thermally averaged cross-section for this process times the DM number density is given by
\beq
n_\chi \langle \sigma v \rangle \sim  
T^3\ \frac{\pi \alpha_W \alpha_d}{T^2}\, .
\label{eq:darkgluon-equilibrium}
\eeq
Comparing this to the Hubble rate $H$ we find that the dark gluons are in equilibrium with the SM at $T \sim M_\chi$ for
\[
\alpha_d \gtrsim \frac{1}{\alpha_W} \, \frac{M_\chi}{M_\text{Planck}}.
\]

Thus even for $\alpha_d$ as small as $10^{-13}$ the dark gluons come to chemical and thermal equilibrium with the SM at temperatures of order $M_\chi$. However, when the universe cools below $T \sim M_\chi$ dark matter becomes non-relativistic and its number density drops exponentially. Then the rate for the process in Figure~\ref{fig:darkgluon-compton} becomes%
\footnote{Note that for temperatures below the W mass, the W boson in Figure~\ref{fig:darkgluon-compton} should be replaced by a photon and the $\alpha_W$ in Eq.~\ref{eq:gluonfreezeout} by $\alpha_{em}$.}
\beq
n_\chi \langle \sigma v \rangle \sim
 (M_\chi T)^{3/2} \, e^{-M_\chi/T} \ \ \frac{\pi \alpha_W \alpha_d}{M_\chi^2}\, ,
\label{eq:gluonfreezeout}
\eeq
and the dark gluons decouple at temperatures of about $M_\chi$ for $\alpha_d \sim 10^{-13}$ to about $M_\chi/20$ for $\alpha_d \sim 10^{-3}$. Below the decoupling temperature the dark radiation fluid evolves independently with a  temperature $T_d$ which redshifts as $1/a$.

The temperature of the photon fluid also redshifts as $1/a$ for most of the universe's evolution. However, when massive SM particles become non-relativistic they annihilate into the remaining lighter SM particles which effectively heats up the photons compared to the dark gluons (similarly to what happens to photons and neutrinos after neutrino decoupling~\cite{Kolb:1990vq}). The ratio between the photon and dark gluon temperatures can be easily calculated in the instantaneous decoupling approximation by requiring that the entropy per co-moving volume is conserved independently in each fluid,\beq
\frac{T_d}{T_\gamma} = \left( \frac{g_*^f}{g_*^i} \right)^{1/3},
\label{eq:temperature-ratio}
\eeq
where $g_*^i$ is the number of effective degrees of freedom in the SM plasma at the time of dark gluon decoupling and $g_*^f$ the number of effective degrees of freedom at any later time.

The CMB places strong constraints on the energy density in relativistic particles at the time of recombination. This constraint is usually presented in terms of the number of effective neutrino species, $N_{eff}$. The contribution of the dark gluons to $N_{eff}$ is given by
\beq
\Delta N_{eff} = \frac{8}{7} \, (N^2-1)  \left( T_d/T_\nu \right)^4,
\eeq
where the $N^2-1$ is the number of generators of \SUN and $T_\nu$ is the neutrino temperature. The ratio can be calculated using Eq.~\ref{eq:temperature-ratio} right at neutrino decoupling when neutrino and photon temperatures are still the same. Assuming that the decoupling between dark gluons and the SM happens at temperatures around $50 \, \text{GeV}$ one finds
\[
\Delta N_{eff} = 0.07 \, (N^2-1).
\]

The strongest constrain on $N_{eff}$ comes from the 2015 Planck data~\cite{Planck:2015xua}, which found $N_{eff} = 3. 15 \pm 0.46$ at $95 \% $ confidence to be compared with the SM prediction~\cite{Mangano:2001iu} $N_{eff}^{SM}=3.046$. We see that this rules out $N\geq 4$ and that $N=3$ is within the $2 \sigma$ allowed range.
However, we note that the Planck analysis assumes the $\Lambda$CDM model with one additional parameter, $N_{eff}$. This limit could potentially be relaxed in our scenario where there are important differences as we will now discuss. 

The effects of dark radiation can be divided into so-called background effects and perturbation effects. Background effects are due to a change in the average energy density in relativistic degrees of freedom and are not sensitive to any other properties of the dark radiation fluid. The largest background effect of extra radiation from relativistic degrees of freedom is to change the redshift of matter-radiation equality $z_{eq}$. Since $z_{eq}$ is very well measured a fit to the data is forced to maintain the redshift of matter-radiation equality by simultaneously increasing the dark matter density~(see e.g.~\cite{Hu:1998tk,Bowen:2001in,Bashinsky:2003tk}). This change in the matter density in turn requires a change in the Hubble parameter today $H_0$, in order to keep $\Omega_m$ fixed (the ratio between the matter density and the critical density). Thus we see that a fit to the CMB data alone has an approximate flat direction in which an increase in $N_{eff}$ can be compensated for by simultaneous increases in $\rho_{DM}$ and $H_0$.

The perturbation effects are due to perturbations in the dark radiation fluid and thus sensitive to properties of dark radiation. In particular there are two additional parameters which distinguish different types of dark radiation, see e.g.~\cite{Diamanti:2012tg}: the effective sound speed $c_\text{eff}^2$ and the viscosity speed $c_\text{vis}^2$. The dark gluons are a relativistic fluid and have $c_\text{eff}^2 = 1/3$, the same as for neutrinos. However---unlike neutrinos---the dark gluons have self-interactions which come from the non-Abelian gauge kinetic terms. If the rate of dark gluon-gluon scattering is large compared to the Hubble parameter then the dark gluons are well described as an ideal fluid instead of a free-streaming fluid as for neutrinos. For an ideal fluid one has $c_\text{vis}^2 = 0$ instead of $1/3$ as for neutrinos. The interaction rate between dark gluons is approximately given by
\beq
\tau^{-1} \sim \alpha_d^2 T_d,
\eeq
and one sees that as long as $\alpha_d \gtrsim 10^{-13}$ this rate is larger than $H$ during recombination. Thus on the time scale set by Hubble the dark gluons behave as a perfect fluid. Qualitatively, the interactions reduce the damping of overdensities from relativistic particles streaming out of gravitational potential wells. This has the effect that the CMB peaks are not as suppressed as they would be in the case of additional free-streaming dark radiation.

The Planck Collaboration has performed a fit for $c_\text{eff}^2$ and $c_\text{vis}^2$ with $N_{eff}$ fixed to the SM value $3.046$, i.e. no additional dark radiation, and found that the parameters were in perfect agreement with the expected value for neutrinos, $c_\text{eff}^2=c_\text{vis}^2=1/3$. However, Planck has not yet performed a fit for additional radiation $\triangle N_{eff}$ which is allowed to have non-standard values for $c_\text{eff}$ and $c_\text{vis}$. In particular, if future CMB experiments find evidence for a non-zero $\triangle N_{eff}$ then measuring $c_\text{vis}$ of this extra component would allow one to distinguish between dark gluons and free-streaming dark radiation, like dark photons or sterile neutrinos~\cite{Friedland:2007vv,Diamanti:2012tg}.

\section{Dark matter-dark gluon interactions and large scale structure}

\begin{figure}[ht]
\centering
\includegraphics[width=0.4\textwidth]{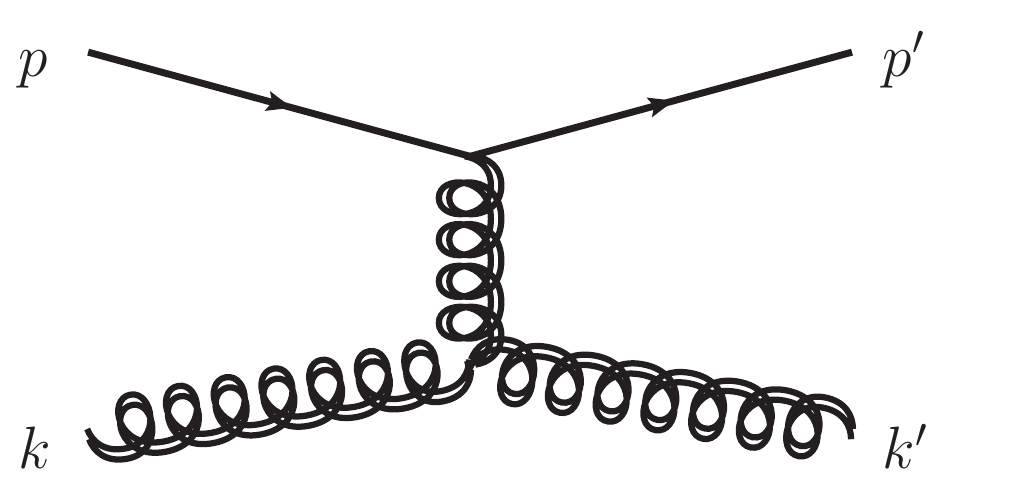}
\caption{t-channel scattering of dark matter with dark gluons.}
\label{fig:t-channel}
\end{figure}

In this Section we study how interactions between dark gluons and DM affect the linear evolution of DM overdensities. The interactions have two important effects. One is the transfer of kinetic energy (i.e. temperature) from the DR to the DM. 
The other is that DM particles moving through the DR fluid experience a drag force.
This drag slows the growth of large scale structures through gravitational clustering. Both rates can be computed by considering scattering of dark gluons with DM particles. 
The process is a generalization of Compton scattering to the non-Abelian case. The most important new feature is that scattering is dominated by the t-channel diagram shown in Figure~\ref{fig:t-channel} which is divergent for small angle scattering and which only exists for non-Abelian gauge bosons. As we will see, this interaction leads to interesting signatures in large scale structure which distinguish our scenario from other models with interactions between dark radiation (DR) and DM (see e.g. \cite{Diamanti:2012tg,Wilkinson:2014ksa}). In particular, for $\alpha_d \simlt 10^{-8.5}$ our  model predicts a smooth suppression of the matter power spectrum at all scales which could resolve the conflict between the indirect measurement of $\sigma_8$ from Planck \cite{Planck:2015xua} and the direct measurement from BOSS \cite{Beutler:2014yhv}.

To simplify the calculation of the energy transfer rate we consider the limit in which the DM temperature is negligible so that we can take the DM particles to be at rest. We then compute the rate of energy transfer to the DM from scattering~\cite{McDermott:2010pa,Dvorkin:2013cea} with a thermal bath of dark gluons. As in the well-known case of Coulomb scattering the cross section is dominated by small-angle forward scattering. To significantly impact the energy of a massive DM particle many collisions with the gluons are required. The collisions are uncorrelated so that the resulting momentum of the DM particle performs a random walk with
$$ E=\frac{p^2}{2M_\chi} \simeq \frac{N}{2M_\chi} (\delta p)^2 \simeq \frac1{2M_\chi} \sum (\delta p)^2\,. $$
Here $\delta p$ is the typical momentum transfer in a single collision and $N$ is the number of collisions. After many such random scatters the resulting DM particle distribution is thermal, but not necessarily with the temperature of the gluon bath. The DM temperature depends on the relative size of the Hubble expansion rate and the energy transfer rate.

The rate of energy transfer is calculated by averaging the energy transfer per collision over the initial Bose-Einstein distribution of the dark gluons
\bea
\dot{E}=a\! \int\! \frac{d^3k}{(2\pi)^3}&& \!\!\!\!2 (N^2\!-\!1) f(k) \times  \\ 
&&\frac{1}{4 E_p k}\int\!  \frac{d^3k'}{(2\pi)^3 2k'} \frac{d^3p'}{(2\pi)^3 2E'_p}
(2\pi)^4 \delta(p\!+\!k\!-\!p'\!-\!k') |\overline{M}|^2 (E'_p\!-\!E_p) (1\!+\!f(k')) \nonumber \,.
\eea
Here---and for the remainder of this Section---time derivatives are taken with respect to conformal time which is the origin of the scale factor $a$ on the right hand side. Also, $f(k)=1/(\exp(k/T_{dr}-1)$ is the gluon thermal distribution function and it is multiplied by $2(N^2-1)$ for the spin and color of the initial gluon. The $1+f(k')$ final state factor accounts for stimulated emission. Finally, the color and spin summed and averaged matrix element (keeping only the t-channel) is given by
$$|\overline{M}|^2 = \frac12\ 4 g_d^4\, \frac{(s-M_\chi^2)(M_\chi^2-u)}{t^2} \ ,$$
where the $1/2$ is the disappointingly boring color factor of the t-channel diagram and $s$, $t$, and $u$ are the Mandelstam variables.
The integrals are straightforward to evaluate for $p=(M_\chi,\vec0)$ and give 
\beq
\dot{E}= a (N^2\!-\!1)\, \frac{\pi}3\,  \alpha_d^2 \log(\alpha_d^{-1})\ \frac{T_{dr}^3}{M_\chi}=
a \frac5\pi \,  \alpha_d^2 \log(\alpha_d^{-1})\ \frac{\rho_{dr}}{T_{dr} M_\chi}\ .
\label{Eq:energy_rate}
\eeq
One subtlety one encounters is a logarithmic divergence at small $t$ which stems from the long range of the interaction mediated by a massless gluon. In the thermal plasma of gluons the interaction range is made finite by screening.  This effect can be parametrized by including a Debye mass for the gluon, $m^2_{Debye}\sim g^2_d T_{dr}^2$ \cite{Arnold:1995bh}. With the Debye mass the logarithmic divergence becomes the $\log(\alpha_d^{-1})$ in Eq.~\ref{Eq:energy_rate}. This is analogous to the ubiquitous ``Coulomb logarithm" in plasma physics.

Using $T=2/3\, E$ we obtain an equation for the evolution of dark matter temperature \cite{Ma:1995ey}
\beq
\dot T_\chi = -2 \,\frac{\dot{a}}{a}\, T_\chi +  a (N^2\!-\!1) \frac{2\pi}{9}\alpha_d^2 \log(\alpha_d^{-1}) \frac{T_{dr}^2}{M_\chi} (T_{dr}-T_\chi) \ ,
\eeq
where the $T_{dr}-T_\chi$ factor generalizes the energy transfer rate in Eq.~\ref{Eq:energy_rate} to the case in which the thermal motion of the DM is not negligible. To understand the possible solutions to this equation one compares the size of the temperature transfer term with the Hubble redshift term $\dot{a}/a\, T_\chi \sim a \, T_\chi T^2/M_{Planck}$ (during radiation domination). For large temperature transfer rates DM and DR are kept in thermal equilibrium by the interactions and $T_\chi=T_{dr}$. Note that both terms scale with the cube of the temperature, thus if DM and DR are in equilibrium at one temperature they will be in equilibrium throughout radiation domination. 

\begin{figure}[ht]
\centering
\includegraphics[width=0.6\textwidth]{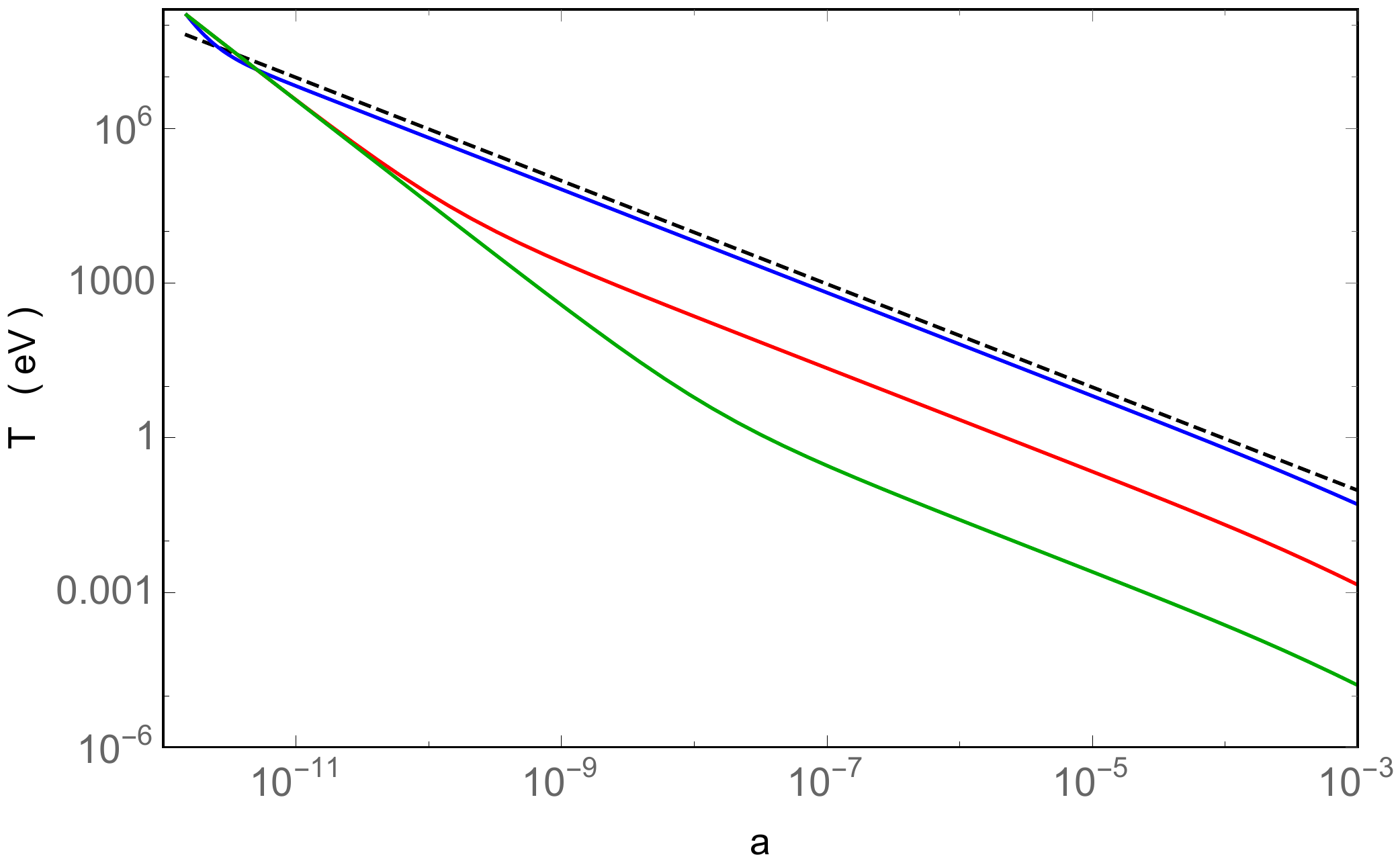}
\caption{Dark sector temperatures as a function of scale factor $a$. Shown are the dark gluon temperature (black dashed), and DM temperatures for three representative values of $\alpha_d=10^{-8}$ (upper, blue), $\alpha_d=10^{-9}$ (middle, red), and $\alpha_d=10^{-10}$ (lower, green).}
\label{fig:temps}
\end{figure}

For couplings $\alpha_d < 10^{-8}$ and assuming comparable starting temperatures $T_\chi \sim T_{dr}$, the Hubble term initially dominates over the interaction term and the DM temperature plummets relative to the temperature of radiation: $T_\chi \sim a^{-2}$ versus $T_{dr} \sim a^{-1}$. This is the regime where our DM behaves like ordinary cold DM (CDM). However, note that the Hubble term scales proportional to $T_{dr}^2 T_\chi$ whereas the collision term scales like $T_{dr}^3$. Thus for sufficiently small $T_\chi$ the two terms become comparable and the DM temperature switches to scaling proportional to $a^{-1}$, keeping the ratio $T_\chi/T_{dr}$ constant. In Figure~\ref{fig:temps} we show numerical solutions for the DM temperature as a function of scale factor for three representative values of $\alpha_d$.

\begin{figure}[ht]
\centering
\includegraphics[width=0.65\textwidth]{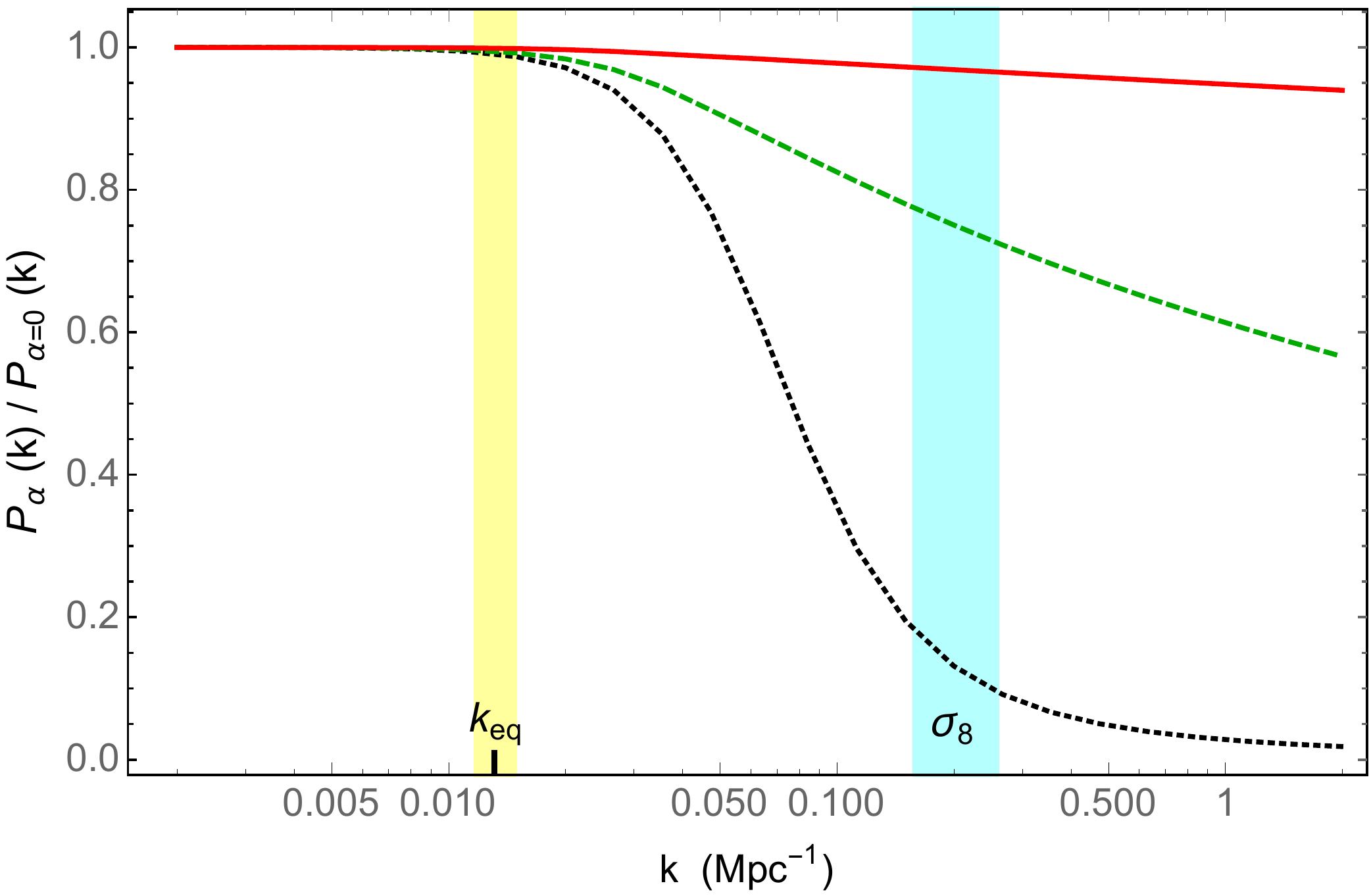}
\caption{Power spectrum including the DM-DR interactions normalized by the power spectrum with interactions turned off. The black dotted curve corresponds to $\alpha_d = 10^{-8}$, the green dashed curve corresponds to $\alpha_d = 10^{-8.5}$ and the red line corresponds to $\alpha_d = 10^{-9}$. The power spectra are defined proportional to $\delta_{DM}^2$ at $a = 10^{-3}$. The vertical yellow band labeled $k_{eq}$ indicates modes which enter the horizon at matter-radiation equality, modes which enter the horizon earlier are to the right (larger $k$). The blue band labeled $\sigma_8$ indicates modes which the observable $\sigma_8$ is most sensitive to.}
\label{fig:power-spectrum}
\end{figure}

For values of $\alpha_d$ for which DM and DR are in thermal equilibrium DM behaves very differently from ordinary CDM. The pressure from the dark gluons prevents the growth of DM overdensities during radiation domination. This can be seen as the sharp drop in the DM power spectrum in Figure~\ref{fig:power-spectrum}. For smaller values of $\alpha_d$ the dark gluons still influence the evolution of DM overdensities, however the effects are more subtle and we employ the formalism of Ma and Bertschinger \cite{Ma:1995ey} to study them. 

Following \cite{Ma:1995ey} we write down the linearized evolution equations for overdensities including the interactions between the DM and DR fluids in conformal Newtonian gauge.
To avoid the complications of solving a full Boltzmann code we work with a simplified scenario. We replace all relativistic energy density in the SM (i.e. neutrinos and photons) with an identical energy density which is made up of only photons. And we replace all matter (dark matter and baryons) with an equivalent energy density of only dark matter. In addition, we approximate by treating the photons as a perfect fluid (zero viscosity). This is only true before recombination but soon after recombination photons contribute only a negligible amount to the energy density and thus to the evolution of DM overdensities. Since our goal is to demonstrate the effect of the interactions on the DM density perturbations we compare our scenario with interactions to the same scenario with $\alpha_d=0$.

In Fourier space the equations for the DM and DR overdensities are
\beq
\begin{aligned}
& \dot \delta_{DM} = -\theta_{DM} + 3 \dot \psi \\
& \dot \theta_{DM}  = - \frac{\dot a}{a} \theta_{DM} + a \tau^{-1}_c (\theta_{DR} - \theta_{DM} ) + k^2 \psi \\
& \dot \delta_{DR}  = - \frac{4}{3} \theta_{DR} + 4 \dot \psi \\ 
& \dot \theta_{DR}  = k^2 \frac{\delta_{DR}}{4} + k^2 \psi + \frac{3}{4}\frac{\rho_{DM}}{\rho_{DR}} a \tau^{-1}_c (\theta_{DM} - \theta_{DR}) \\
& \dot \delta_\gamma  = - \frac{4}{3} \theta_\gamma + 4 \dot \psi \\
& \dot \theta_\gamma  = \frac{1}{4} k^2 \delta_\gamma + k^2 \psi \\
& k^2 \psi + 3 \frac{\dot a}{a} \left( \dot \psi + \frac{\dot a}{a} \, \psi \right) =  - \frac{a^2}{2 M_{Pl}^2} \sum_i \rho_i \delta_i \, ,
\end{aligned}
\eeq
where the dots represent derivatives with respect to conformal time, $\rho_X$ is the average energy density of fluid $X$ and $\delta_X$ and $\theta_X$ are related to the overdensity and velocity divergence in fluid $X$. We have also set the two metric perturbations $\psi$ and $\phi$ equal because we are treating the photons and dark radiation as ideal fluids (no anisotropic stress) and did not include neutrinos which have sizable anisotropic stresses. The interaction between dark matter and dark radiation is encoded in the momentum transfer rate $\tau^{-1}_c$~\cite{Dvorkin:2013cea}. It is defined as the change in momentum $\dot{\vec{p}}_\chi = - a \tau^{-1}_c\ \vec p_\chi$ which a DM particle with momentum $\vec{P}$ experiences due to friction as it is moving through the dark gluon fluid. Microscopically, the friction arises from collisions between DM particles and dark gluons and to compute it we evaluate 
\beq
\dot{\vec{p}}=a\! \int\! \frac{d^3k}{(2\pi)^3}f(k) \frac{1}{4 E_p k}\int\!  \frac{d^3k'}{(2\pi)^3 2k'} \frac{d^3p'}{(2\pi)^3 2E'_p}
(2\pi)^4 \delta(p\!+\!k\!-\!p'\!-\!k') |M|^2 (\vec{p}{\,'}\!-\!\vec{p}\,) (1\!+\!f(k')) \ ,
\eeq
where now the initial DM momentum $\vec{p}$ is non-zero and we expand to first order in $p/M_\chi$. Employing the same approximations as for the energy transfer rate we obtain
\beq
\tau^{-1}_c = (N^2\!-\!1) \frac{\pi}{9} \alpha_d^2 \log\alpha_d^{-1} \, \frac{T_{dr}^2}{M_\chi}
\label{eq:velocity-rate}
\eeq

We integrate the equations for the overdensities from $a=10^{-7}$, when all modes of interest are well outside the horizon, until $a=10^{-3}$. We use initial conditions corresponding to adiabatic perturbations:
\beq
\begin{aligned}
\delta_\gamma = \delta_{DR} = \frac{4}{3} \delta_{DM} = - 2 \psi = C(k) \, ,
\end{aligned}
\label{eq:initial_conditions}
\eeq
where the initial perturbations $C(k) \sim k^{-3/2}$ are determined by the physics of inflation. We define the DM power spectrum equal to the square of the perturbations, $P(k)\equiv \, \delta_{DM}^2$, at scale factor $a=10^{-3}$. To focus only on effects of the coupling between DM and DR we form a ratio where we divide the power spectrum with interactions turned on by the power spectrum with $\alpha_d = 0$. Note that since the equations are linear the initial values for the perturbations, $C(k)$, drop out in the ratio.

The ratios of power spectra for different values of $\alpha_d$ are plotted in Figure~\ref{fig:power-spectrum}. 
For the plot we chose the number of colors $N=2$ and $M_\chi = 1.2\,$TeV. One sees that for $\alpha_d = 10^{-8}$ the power spectrum is strongly suppressed for modes which entered the horizon before matter-radiation equality. These are modes with $k > k_{eq} \sim 0.015 \, \text{Mpc}^{-1}$. This should be expected because in this case the DM is in equilibrium with the DR bath throughout radiation domination. For the smaller values of $\alpha_d = 10^{-8.5}$ and $10^{-9}$ the power spectrum is less affected with modes which entered the horizon earlier (larger $k$) suppressed more that those which entered later. Modes which enter the horizon after matter radiation equality are not suppressed for any of the couplings plotted. The light blue vertical band indicates the range of modes which the observable $\sigma_8$ is sensitive to ($\sigma_8$ is a measurement of the matter fluctuations in spheres of radius of $8 h^{-1}$ Mpc).

The smooth suppression of power at all scales that we are finding is special to our scenario and stems from the fact that the momentum transfer rate scales with temperature as $T_{dr}^2$, the same scaling as Hubble. Thus it is possible to arrange for the couplings to have a small effect but act over a large range of scales. This should be contrasted with cases where the interactions scale like a higher power of $T_{dr}$ in which case they are important at high scales and have no effect at low scales. In such scenarios the power spectrum has a sharp cutoff at scales of the size of the horizon at the time when the interactions cease to be important, leaving larger scales unaffected and wiping out the smaller scales that entered the horizon at earlier times.

The smooth suppression of power is exciting because it might help resolve two sources of tension in recent experimental results. A fit to the most recent Planck CMB data is used to fix the parameters of $\Lambda$CDM. Using the model, the Planck collaboration predicts $\sigma_8^{\Lambda CDM}=0.829\pm0.014$ (``TT + lowP", 1$\sigma$ errors). This value is about 2$\sigma$ higher than ``direct" measurements of large scale structure ~\cite{Beutler:2014yhv,Battye:2014qga,Planck:2015xua} (for example, gravitational lensing of the CMB as measured by Planck gives $\sigma_8^{lensing}=0.802\pm0.012$). A reduction of the predicted power spectrum  due to DM-DR interactions as, it occurs in our model for $\alpha_d\sim 10^{-9}$, removes this tension between the Planck fit and LSS data. Interestingly, the tension in $\sigma_8$ is currently also driving the Planck fit for $N_{eff}$ and $H_0$ to lower values, because in $\Lambda$CDM larger values for those parameters would correlate with even larger values for $\sigma_8$. Thus after including DM-DR interactions the Planck fit might prefer $N_{eff}> 3$ which would create more room for dark gluons. Furthermore, larger $N_{eff}$ is correlated in the $\Lambda$CDM fit with a larger value for $H_0$ (to keep the position of the acoustic peaks in the CMB fixed~\cite{Hu:1998tk,Bashinsky:2003tk,Bowen:2001in,Planck:2015xua}). This in turn would allow better agreement between the $H_0$ values from Planck and supernova data~\cite{Planck:2015xua}, another area of mild tension in cosmological data. Clearly, a more quantitative analysis of this issue is desirable and requires including the non-Abelian dark radiation in a full Boltzmann code.

\section{Conclusions}

In this work we have studied a new type of dark sector with massless non-Abelian gauge bosons super-weakly coupled to the DM. There are many different possibilities for the DM coupling to the standard model. As an example, we chose our DM particle to transform as \SUw triplets and fundamentals under the dark \SUN gauge group.

Our model has three new parameters, the coupling constant $\alpha_d$, the mass of the DM particles $M_\chi$, and the size of the gauge group $N$. Demanding the correct DM abundance from thermal freeze-out fixes the DM mass in terms of $N$, thus leaving a two dimensional parameter space. Constraints on this parameter space can be derived from several different experiments.

The first set of constraints derives simply from the multiplicity of the dark matter and would even apply if the dark coupling constant were zero. The multiplicity of dark matter affects the usual WIMP searches (direct and indirect detection and collider searches). The effect is simply that dark color multiplicity factors enhance pair production and decrease pair annihilation. Therefore the DM mass required in order to predict the right DM abundance decreases by $\sqrt{2 N}$. It also increases the collider cross-section, placing this type of dark matter within easy reach of the proposed $100$ TeV collider. The decrease in mass and in annihilation cross-section also removes the current tension between thermally produced \SUw triplet dark matter and H.E.S.S data. Despite the decrease in annihilation cross-section, the \SUw triplet DM model annihilation cross-section is within the projected reach for CTA.

For $\alpha_d \simgt 10^{-13}$ the dark gluons thermalize with the SM in the early universe and contribute to dark radiation. The limits placed on $N_{eff}$ from Planck constrain $N$ to be at most $3$. However, the self interactions of the dark gluons leave an imprint in the CMB which is distinct from that of free-streaming fluids like neutrinos or dark photons. This can be used to distinguish between the two types of radiation if future experiments establish the need for a non-standard contribution to $N_{eff}$.

Finally we studied the effects of the interactions between DM and DR on the power spectrum. We found that for $\alpha_d \gtrsim 10^{-8}$ the interactions strongly suppress the power spectrum of modes entering the horizon before matter radiation equality and thus such couplings are ruled out. On the other hand, for $\alpha_d \simlt 10^{-8.5}$ the interactions predict a smooth decrease in the power spectrum, which can potentially solve the discrepancy between Planck and large scale structure data and the discrepancy between Planck and Supernova measurements of $H_0$.  We hope to return to this in future work.

\section{Acknowledgments}
We wish to thank Ed Bertschinger, Chris Brust, Mark Hertzberg, David Kaplan, Lloyd Knox, Michele Papucci, Tracy Slatyer, Jesse Thaler, Matt Walters, Kathryn Zurek,  and especially Andy Cohen and Julien Lesgourgues for useful discussions. MS acknowledges support from the Aspen Center for Physics where some of the early work on this paper was performed. This work was supported by the US Department of Energy Office of Science under Award Number DE-SC-0010025. GMT also acknowledges support from a DOE High Energy Physics Graduate Fellowship.


\bibliographystyle{JHEP}

\bibliography{dark-arxiv}{}

\end{document}